\documentclass[twocolumn,prl,superscriptaddress,showpacs,byrevtex]{revtex4}
\usepackage{epsf,graphicx,amssymb}
\usepackage[usenames]{color}

\begin{document}

\title{Wave turbulence on the surface of a ferrofluid in a horizontal magnetic field}
\author{St\'ephane Dorbolo} 
\affiliation{FRS--FNRS, GRASP, D\'epartement de Physique B5, Universit\'e de Li\`ege, Belgium, EU}
\author{Eric Falcon}
\email[E-mail: ]{eric.falcon@univ-paris-diderot.fr}
\affiliation{Mati\`ere et Syst\`emes Complexes (MSC), Universit\'e Paris Diderot, CNRS -- UMR 7057\\ 10 rue A. Domon \& L. Duquet, 75 013 Paris, France, EU}

\date{\today}

\begin{abstract}  
We report observations of wave turbulence on the surface of a ferrofluid submitted to a magnetic field parallel to the fluid surface. The magnetic wave turbulence shows several differences compared to the normal field case reported recently. The inertial zone of the magnetic wave turbulence regime is notably found to be strongly increased with respect to the normal field case, and to be well described by our theoretical predictions. The dispersion relation of linear waves is also measured and differs from the normal field case due to the absence of the Rosensweig instability.
\end{abstract}

\maketitle
\section{Introduction}
Wave turbulence concerns the dynamical and statistical properties of numerous nonlinearly interacting waves within a dispersive medium. Contrary to hydrodynamical turbulence, out-of-equilibrium solutions for the power spectrum of waves can be analytically computed by the wave turbulence theory in nearly all fields of physics (e.g., oceanic surface or internal waves, elastic waves on a plate, plasma waves, and spin waves) ~\cite{SergeyLivre,ZakharovLivre,Newell11}. This theory assumes strong hypotheses such as weak nonlinearities and infinite size systems. Although the number of wave turbulence experiments has strongly increased in the last decade (see \cite{Falcon10} for a review), they are still lagging behind the theory. New experimental systems are thus expected to check the validity domain of the theory in experiments and to benefit from the wave turbulence theoretical framework. 

For instance, an operator can easily change the dispersion relation of linear waves propagating on the surface of a magnetic fluid by applying a uniform magnetic field \cite{Rosen}. Magnetic wave turbulence can then be achieved on a ferrofluid surface submitted to a vertical field \cite{Boyer08}. When the vertical field is high enough, the so-called Rosensweig instability occurs leading to a hexagonal pattern of peaks on the surface of the ferrofluid \cite{Rosen}. Waves on the surface of a magnetic fluid submitted to a horizontal magnetic field are much less studied. In this geometry, the dispersion relation of linear waves is always monotonous whatever the field intensity and the Rosensweig instability is absent  \cite{Rosen}. To our knowledge, only one experimental test of the dispersion relation has been performed in this configuration \cite{Zelaco69} although striking phenomena occur: strong anisotropy of propagating waves, shifting onsets of the Kelvin-Helmholtz and Rayleigh-Taylor instabilities \cite{Rosen,Zelaco69}, and a new pattern of the magnetic Faraday instability \cite{Bacri94}. 

In this paper, we report observations of wave turbulence on the surface of a ferrofluid submitted to a horizontal magnetic field.  We show that it displays significant qualitative differences with respect to the vertical field case. The most striking one is the extension of the inertial zone of the magnetic wave turbulence regime. Moreover, the magnetic wave turbulence is found to be isotropic (whatever the tangential magnetic field direction). The frequency-power law scalings of the spectra of the magnetic and capillary wave turbulence regimes are also found to depend on the magnetic field applied contrary to the predictions based on non interacting regimes. The dispersion relation of linear waves is also measured and discussed.
 
\begin{figure}[t!]
\includegraphics[height=45mm]{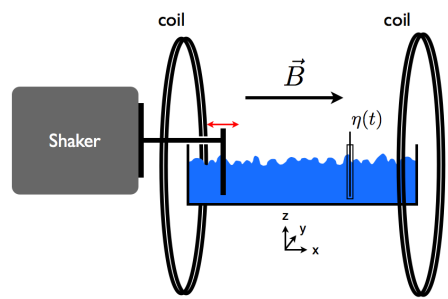}
\caption{(color online). Experimental setup.}
\label{exp}
\end{figure}

\section{Experimental setup}
The experimental setup is shown in Fig.\ \ref{exp}. It consists of a cylindrical container, 120 mm in inner diameter and 40 mm in depth, half filled with a ferrofluid. The ferrofluid used is a ionic aqueous suspension synthesized with 8.5\% by volume of magnetic particles (Fe$_2$O$_3$ ; 7.6 $\pm 0.36$ nm in diameter) \cite{Talbot}. The properties of this magnetic fluid are: density, $\rho=1324$ kg/m$^3$, surface tension $\gamma=59 \times 10^{-3}$ N/m, initial magnetic susceptibility, $\chi_i=0.69$, and magnetic saturation $M_{sat}=16.9\times 10^{3}$ A/m. The container is placed between two coaxial coils, 250 mm (500 mm) in inner (outer) diameter, 70 mm apart as in Fig.\ \ref{exp}. The container and the coil axes are perpendicular, leading to an applied magnetic field parallel to the ferrofluid surface. The orientation of the magnetic field is thus different from that in Ref. \cite{Boyer08}, where the field was normal to the fluid surface. A dc current is supplied to the coils in series by a power supply (50 V/48 A), the coils being cooled with water circulation. The horizontal magnetic induction generated is up to 780 G and is measured by a Hall probe. To make easier the comparison with the normal magnetic field case \cite{Boyer08}, one will normalize in the following the value of the parallel magnetic induction $B$ by the critical value of the normal magnetic induction $B_c$ for the Rosensweig instability, which is found to be $B_c=294\pm2$ G  \cite{Boyer08}. Surface waves are generated on the surface of the ferrofluid by the horizontal motion of a rectangular plunging Teflon wave maker driven by an electromagnetic vibration shaker (see Fig.\ \ref{exp}). The motion of the wave maker is collinear with the direction of the horizontal magnetic field (i.e., in the $x$-axis, see Fig.\ \ref{exp}) unless otherwise stated. The shaker is either driven sinusoidally  in a range 5 $\leq f \leq$ 100 Hz to measure the dispersion relation of linear waves, or with a random vibration between 1 and 6 Hz for wave turbulence experiments. This latter forcing generates stochastic waves on the surface of the ferrofluid \cite{Boyer08}. The amplitude of the surface waves $\eta(t)$ at a given location is measured by a capacitive wire gauge plunging perpendicularly to the fluid at rest (see Fig.\ \ref{exp}) \cite{Falcon07}. The signal $\eta(t)$ is analogically low-pass filtered at 1 kHz and is recorded for 800 s using an acquisition card with a 4 kHz sampling rate.

\section{Dispersion relation and crossover frequencies}
In the deep fluid approximation, the dispersion relation of linear inviscid surface waves on the surface of a magnetic fluid submitted to a horizontal magnetic induction $B$ collinear to the vibration direction (i.e. in the $x$-axis - see Fig. 1) reads \cite{Zelaco69,Rosen}
\begin{equation}
\omega^2=gk+\frac{F[\chi]}{\rho\mu_0}B^2k_x^2+\frac{\gamma}{\rho}k^3
\label{rdtheo}
\end{equation}
with $\omega\equiv 2\pi f$, $f$ the frequency, $k \equiv \sqrt{k_x^2+k_y^2} \equiv 2\pi /\lambda$, $\lambda$ the wavelength, $g=9.81$ m/s$^2$ the acceleration of the gravity,  $\mu_0=4\pi \times 10^{-7}$ H/m the magnetic permeability of the vacuum, and $F[\chi]\equiv \chi^2/[(2+\chi)(1+\chi)]$. $\chi$ is the magnetic susceptibility of the ferrofluid which depends on the applied magnetic field, $H$, through Langevin's classical theory \cite{Abou97} 
\begin{equation}
\chi(H)=\frac{M_{sat}}{H}\mathcal{L}\left(\frac{3\chi_iH}{M_{sat}}\right)
\label{aimantation}
\end{equation}
where $\mathcal{L}(x)\equiv \coth{(x)}-1/x$, and thus on the magnetic induction, $B$, through an implicit equation, since
 \begin{equation}
B=\mu_0(1+\chi)H
\label{BvsH}
\end{equation}
Note that if the horizontal magnetic field $B$ is applied in the $y$-axis direction normal to the direction of vibration, the second term in Eq.\ (\ref{rdtheo}) vanishes  \cite{Rosen}. Moreover, the difference with the case where the magnetic induction is normal to the fluid surface (i.e., in the $z$-axis) is the sign of the second term of the right-hand side of Eq.\ (\ref{rdtheo}). This prevents the Rosensweig instability \cite{Rosen}, and the dispersion relation of Eq.\ (\ref{rdtheo}) is then monotonous. This latter is dominated by the gravity waves at small $k$, and by the capillary waves at large $k$ whatever the value of $B$. Magnetic waves are assumed to occur when the quadratic term dominates the linear and the cubic term  \cite{Boyer08}. This arises when $F[\chi]B^2 > \mu_0\sqrt{\rho g \gamma}$ that is using Eqs.\ (\ref{aimantation}) and (\ref{BvsH}) and the ferrofluid properties when $B/B_c>0.65$ \cite{Boyer08}.

Let us now compute the magnetic field dependence of the crossover frequencies between gravity and magnetic waves, magnetic and capillary waves, and gravity and capillary waves for the case of an horizontal magnetic field. Crossovers are derived by balancing the terms of Eq.\ (\ref{rdtheo}) each to each. In order to indicate the difference between the case when the magnetic field is normal or parallel to the fluid surface, the crossover frequencies are denoted by the subscript ${\perp}$ and ${\parallel}$ respectively. For the gravity-capillary transition, one balances the first and the third terms of the right-hand side of Eq.\ (\ref{rdtheo}), that is $gk_{gc}=(\gamma/\rho)k_{gc}^3$, thus for $k_{gc}=\sqrt{\rho g/\gamma}$ which substituted into Eq.\ (\ref{rdtheo}) gives
\begin{equation}
\omega^2_{gc_{\parallel}}=\omega^2_{gc_{\perp}}+\frac{2gF[\chi]B^2}{\mu_0\gamma} {\rm \ for \ \,Ê} F[\chi]B^2 < \mu_0\sqrt{\rho g \gamma} {\rm \ \ , }
\label{fgc}
\end{equation}
with $\omega^2_{gc_{\perp}}=2\sqrt{g^3\rho/\gamma}-gF[\chi]B^2/(\mu_0\gamma)$ the value found for the normal field case \cite{Boyer08}. Similarly, by balancing the first and second terms, the gravity-magnetic crossover frequency reads 
\begin{equation}
\omega^2_{gm_{\parallel}}=\omega^2_{gm_{\perp}}+\frac{2\rho\mu_0g^2}{F[\chi]B^{2}} {\rm \ for \ \, } F[\chi]B^2 > \mu_0\sqrt{\rho g \gamma} {\rm \ \ , }
\label{fgm}
\end{equation} 
with $\omega^2_{gm_{\perp}}=\left[\mu_0\rho g/F[\chi]B^{2}\right]^3\gamma /\rho$. Finally, by balancing the second and third terms, the magneto-capillary crossover frequency reads
\begin{equation}
\omega^2_{mc_{\parallel}}=\omega^2_{mc_{\perp}} + \frac{2}{\rho\gamma^2}\left[\frac{F[\chi]B^2}{\mu_0}\right]^3 {\rm \ for  \ \, } F[\chi]B^2 > \mu_0\sqrt{\rho g \gamma} {\rm \ \ , }
\label{fmc}
\end{equation}
with $\omega^2_{mc_{\perp}}=gF[\chi]B^2/(\mu_0\gamma)$. 
The crossover frequencies for the horizontal magnetic field configuration (denoted by the subscript ${\parallel}$) have thus an additional term [second term of the right-hand side of Eqs.\ (\ref{fgc}--\ref{fmc})] with respect to the ones found in Ref.\ \cite{Boyer08} for the normal field case (denoted by the subscript ${\perp}$). One can also compute a triple point in the ($\omega$ , $B$) space where the three wave domains (gravity, magnetic, and capillary) coexist. It can be derived by balancing the three terms of Eq.\ (\ref{rdtheo}), which leads to \begin{equation}
f_{t_{\parallel}}=\sqrt{3}f_{t_{\perp}}  {\rm \ \ and  \ \ } B_{t_{\parallel}}=B_{t_{\perp}}  {\rm \ \ , }
\label{ft}
\end{equation}
with $2\pi f_{t_{\perp}}=(g^3\rho/\gamma)^{1/4}$ and $B_{t_{\perp}}^2= \mu_0\sqrt{\rho g \gamma}/F[\chi(H_t)]$. The frequency of the triple point is thus increased in the parallel-field configuration with respect to the normal-field one, whereas the magnetic field at this point is independent of the orientation of the field. By using the properties of our ferrofluid, one thus predicts a triple point located at $f_{t_{\parallel}}=18.7$ Hz and $B_t/B_c=0.65$. 

\begin{figure}[t!]
\includegraphics[height=60mm]{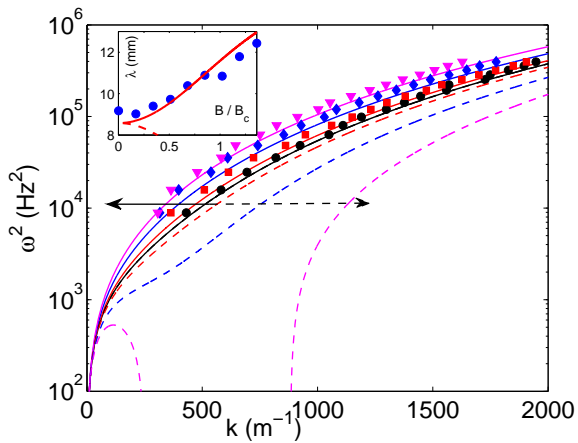}
\caption{(color online). Dispersion relation $\omega^2$ as a function of $k$ for different horizontal magnetic fields $B/B_c=$ ($\bullet$) 0,  ($\blacksquare$) 0.39, ($\blacklozenge$), 0.76, and ($\blacktriangledown$) 1.13. Solid lines correspond to Eq.\ (\ref{rdtheo}) for each value of $B$ (from bottom to top). Dashed lines correspond to the dispersion relation when $B$ is normal to the fluid surface for the same values of $B$ (from top to bottom). Arrows show that, for a fixed $f$, $k$ decreases (resp. increases) with $B$ for the parallel-field (resp. normal-field) case. Inset: $\lambda$ as a function of $B/B_c$ for $f=25$ Hz. Solid line is from Eq.\ (\ref{rdtheo}). $B_c=$294 G.}
\label{disper}
\end{figure}

\section{Experimental results}
\subsection{Dispersion relation}
Let us first measure the dispersion relation of linear waves on the surface of a ferrofluid submitted to a horizontal magnetic field.  The wave maker is driven sinusoidally at a fixed frequency $f$. The wavelength $\lambda$ of the surface waves is measured by detecting the wave crests by using a camera. For a fixed $f$, $\lambda$ is measured for different values of the horizontal magnetic field $B$. As shown in the inset of Fig.\ \ref{disper}, $\lambda$ is of the order of a centimeter and is found to increases with $B$. One also measures the dispersion relation $\omega^2$ vs $k$ by varying $f$ for four values of $B$ as shown in the main part of Fig.\ \ref{disper}. The solid lines are theoretical curves computed from Eq.\ (\ref{rdtheo}) and show as expected a rough agreement with the data with no adjustable parameter. The theoretical dispersion relations for the normal magnetic field case are also computed and displayed in dashed lines in Fig.\ \ref{disper} (for the same values of $B$) for comparison. The parallel-field case shows two main differences from the normal-field case: (i) the dispersion relation is monotonous underlying the absence of the Rosensweig (normal-field) instability, and (ii) the wavelength $\lambda$ increases with $B$ at a fixed frequency contrary to the normal-field case (see arrows in Fig.\ \ref{disper}).  As shown below, different nonlinear wave behaviors will also occur with respect to the magnetic field orientation.

\begin{figure}[t!]
\includegraphics[height=60mm]{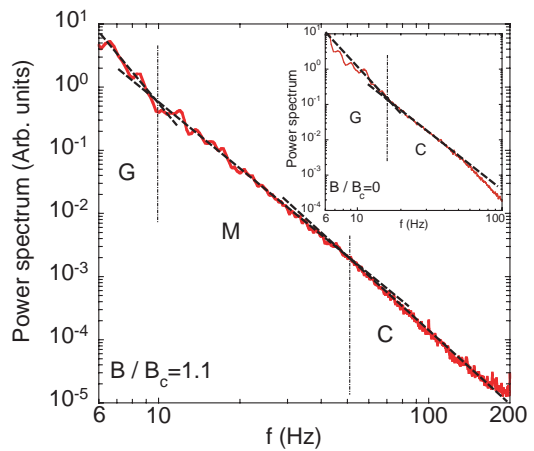}
\caption{(color online). Power spectrum of wave amplitude for two values of $B$. Inset: $B/B_c=0$ showing  gravity (G), and capillary (C) wave turbulence regimes. Solid lines have slopes of -4.5 and -3.2. Crossover: $f_{gc}=$ 15.8 Hz. Main: $B/B_c=1.1$ showing gravity (G), magnetic (M) and capillary (C) wave turbulence regime. Solid lines have slopes of -5.2, -3.5 and -3.9. Crossovers: $f_{gm}=$ 9.8 Hz and $f_{mc}=$ 51.4 Hz. Forcing parameters: 1 - 6 Hz. $B_c= 294$ G.}
\label{fig02}
\end{figure}

\subsection{Wave turbulence power spectrum}
Let us now focus on nonlinear surface waves in a turbulent regime. The wave maker is now driven by a random vibration both in amplitude and frequency in the range from 1 to 6 Hz. This leads to the generation of surface waves of random amplitudes that mix each other erratically. The power spectrum of the wave amplitude is shown in Fig.\ \ref{fig02} for different horizontal magnetic inductions applied $B/B_c$. 

For $B/B_c=0$ (see inset of Fig.\ \ref{fig02}), it displays similar results than those found with a usual fluid \cite{Falcon07} (see also \cite{Boyer08}): a cross-over between gravity and capillary regimes is observed near $f_{gc} \simeq 17$ Hz separating two power laws. These power laws are a signature of gravity and capillary wave turbulence regimes, the end of the capillary power-law being related to the dissipation. The theoretical value of the crossover is $f_{gc}=\frac{1}{2\pi}\sqrt{2g/l_c}\simeq 15.2$ Hz where $l_c=\sqrt{\gamma/(\rho g)}$ \cite{Falcon07}. However, it has been shown that this value  slightly depends on the forcing parameters \cite{Falcon07}. The slope in the gravity and capillary regimes are found to be -4.5 and -3.2, respectively. These values are in pretty good agreement with the expected values for the gravity ($f^{-4}$ \cite{Zakharov67Grav}) and capillary ($f^{-17/6}$ \cite{Zakharov67Cap}) wave turbulence regimes. Note that for the gravity regime, the exponent is known to depend on the forcing parameters \cite{Falcon07}.

For $B/B_c=1.1$ (see Fig.\ \ref{fig02}), one observes three wave turbulence regimes and two crossover frequencies separating these regimes. The gravity regime is roughly observed in Fig. \ref{fig02} below 10 Hz, the magnetic one between 10 and 50 Hz and the capillary regime beyond 50 Hz. Despite these small inertial ranges, the power spectrum is well fitted by different frequency-power laws for the gravity, magnetic, and capillary regimes of exponents -5.2, -3.5 and -3.9 respectively.

\begin{figure}[t!]
\includegraphics[height=60mm]{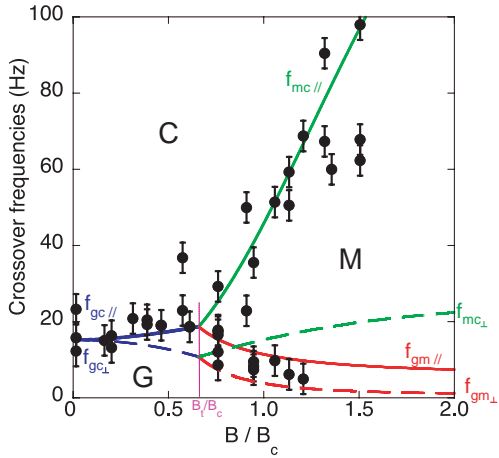} 
\caption{(color online). Crossover frequencies $f_{gc}$, $f_{gm}$ and $f_{mc}$  ($\bullet$) as a function of the horizontal magnetic field $B/B_c$. Solid lines $f_{gc_{\parallel}}$ (middle blue line), $f_{gm_{\parallel}}$ (lower red line), and $f_{mc_{\parallel}}$ (upper green line) are from Eqs.\ (\ref{fgc}), (\ref{fgm}) and (\ref{fmc}) respectively. The triple point ($f_{t_{\parallel}}=18.7$ Hz , $B_t/B_c=0.65$) is from Eq.\ (\ref{ft}). Dashed lines correspond to the theoretical crossovers $f_{gc_{\perp}}$, $f_{gm_{\perp}}$ and $f_{mc_{\perp}}$for the normal field case \cite{Boyer08}. Forcing parameters: 1 - 6 Hz. $B_c= 294$ G.}
\label{fig04}
\end{figure}

The crossovers between gravity and capillary regimes $f_{gc}$, between gravity and magnetic regimes $f_{gm}$, and between magnetic and capillary regimes $f_{mc}$ have been measured for different horizontal magnetic fields applied, as shown in Fig. \ref{fig04}. The solid lines are the prediction from Eqs.\ (\ref{fgc}), (\ref{fgm}) and (\ref{fmc}) that correspond to $f_{gc_{\parallel}}(B)$ (middle blue line), $f_{gm_{\parallel}}(B)$ (lower red line), and $f_{mc_{\parallel}}(B)$ (upper green line), respectively. A rough agreement is obtained between the data and the above predictions with no adjustable parameter: $f_{gc}$ increases with $B$ up to the onset of magnetic waves at $B/B_c\gtrsim 0.65$. This onset is also in good agreement with its predicted value $F[\chi]B^2 > \mu_0\sqrt{\rho g \gamma}$ (see above). For higher $B/B_c$, $f_{mc}$ is found to strongly increase with $B$, whereas $f_{gm}$ slightly decreases. The triple point coexistence of the three regimes of wave turbulence is found at $f_{t}\simeq 19$ Hz and $B_t/B_c\simeq 0.65$ and is also in good agreement with Eq.\ (\ref{ft}). The dependence of these crossover frequencies on the parallel magnetic field is strongly different from the ones found when the magnetic field is normal to the surface of fluid \cite{Boyer08}. Indeed, the dashed lines shown in Fig. \ref{fig04} represent the theoretical crossover frequencies for the normal-field case, and leads to a qualitatively different behavior for $f_{gc_{\perp}}$  and $f_{mc_{\perp}}$ (inversion of curvature) and a quantitatively one for $f_{gm_{\perp}}$. Notably, the inertial zone of the magnetic wave turbulence is strongly extended for a horizontal magnetic field. Note that when the direction of the horizontal magnetic field is applied normally to the vibration direction (i.e., in the $y$-axis, see Fig.\ \ref{exp}), the wave turbulence results are not significantly changed (for our range of $B$) with respect to the case where $B$ is collinear to the vibration (i.e., in the $x$-axis). Consequently, this underlies the isotropy of the observed magnetic wave turbulence, although linear waves are strongly affected by the orientation of the tangential field $B$ (see Eq.\ \ref{rdtheo}).

\begin{figure}[t!]
\includegraphics[height=60mm]{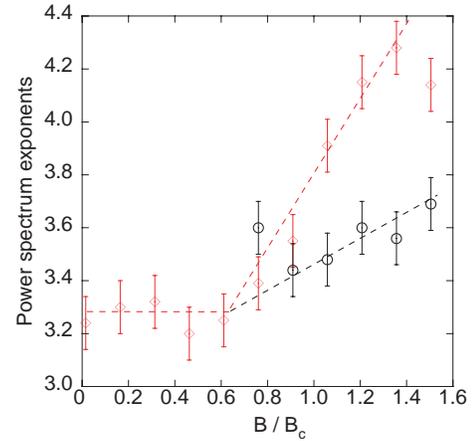}
\caption{(color online). Frequency-power-law exponents (absolute value) of the power spectrum as a function of $B / B_c$ for the magnetic (black circle) and capillary (red diamond) wave turbulence regimes. Dashed lines are a guide to the eye. Forcing parameters: 1 - 6 Hz. $B_c= 294$ G.}
\label{fig03}
\end{figure}

Finally, the exponents of the frequency-power-law spectrum of the magnetic and capillary wave turbulence regimes are shown in Fig.\ \ref{fig03} for different $B$ applied.  The exponent of the gravity wave regime is not reported here, because it occurs on a frequency range too small. For small $B$, capillary waves are dominant and the exponent of the capillary regime is roughly found constant $\simeq 3.3$ (see $\diamond$-symbol) in rough agreement with the capillary predictions \cite{Zakharov67Cap}. When $B/B_c \geq 0.65$, magnetic waves occur (see Fig.\ \ref{fig04}). The exponent of the magnetic regime is then found to increase with $B$ (see $\circ$-symbol in Fig.\ \ref{fig03}) and departs from the value of 3, the one predicted for the magnetic exponent \cite{Boyer08}. Moreover, the evolution of the third power-law exponent (that should correspond to the capillary regime) also increases with $B$ (see $\diamond$-symbol for $B/B_c \geq 0.65$) and departs from the capillary exponent ($\simeq 3.3$) found with no magnetic field. Surprisingly, the magnetic field dependence of both magnetic and capillary exponents is similar to the one reported for the normal-field case \cite{Boyer08} and cannot be thus ascribed to the Rosensweig instability as suggested in \cite{Boyer08}. This cannot be also ascribed to a possible anisotropy of nonlinear waves, since similar wave turbulence results are obtained when the horizontal field is normal or collinear to the vibration direction (see above). The origin of the dependence of these exponents with $B$ could be due to a possible  interaction between the capillary and magnetic wave turbulence regimes, the exponents of each regime being predicted assuming non interacting regimes. This possible origin could be checked by working at much higher values of $B$ and thus deserves further study.

\section{Conclusion}
We have observed wave turbulence on the surface of a magnetic fluid submitted to a magnetic field parallel to the fluid surface. It displays several differences with respect to the normal-field configuration. The most striking one is the extension of the inertial zone of the magnetic wave turbulence regime. The magnetic wave turbulence is also found isotropic, contrary to linear waves that depend on the horizontal magnetic field orientation. Moreover, both frequency-power-law exponents of the power spectrum of the magnetic and capillary wave turbulence are found to depend on the magnetic field applied. This suggests that both regimes interact with each other when magnetic waves occur.

\section{Acknowledgments}
\begin{acknowledgments}
SD thanks FNRS for financial support and University Paris Diderot for their invited researcher funding.  This work has been supported by ANR Turbonde BLAN07-3-197846. This work has been also supported by ``Wallonie-Bruxelles International'' and ``Fonds de la Recherche Scientifique du Minist\`ere Fran\c {c}ais des Affaires Etrang\`eres et Europ\'eennes'', and by ``Minist\`ere de l'Enseignement sup\'erieur et de la Recherche'' in the frame of Partnership Hubert Curien (PHC) Tournesol Fr 22486ZK.
\end{acknowledgments}


\end{document}